\documentclass[twocolumn,aps,pra,showpacs,amsmath,amssymb,superscriptaddress]{revtex4}

\usepackage[usenames]{color}
\usepackage[dvips]{graphicx}

\def\<{\langle}
\def\>{\rangle}
\graphicspath{{converted_graphics/}}

\begin{document}

\title{Three-body Physics in Strongly Correlated Spinor Condensates}

\author{V. E. Colussi}
\affiliation{JILA and Department of Physics, University of Colorado, Boulder, Colorado 80309-0440, USA}
\author{Chris H. Greene}
\affiliation{Department of Physics, Purdue University, West Lafayette, Indiana 47907-2036, USA}
\author{J. P. D'Incao}
\affiliation{JILA and Department of Physics, University of Colorado, Boulder, Colorado 80309-0440, USA}

\begin{abstract}
Spinor condensates have proven to be a rich area for probing many-body phenomena richer than that of an ultracold gas consisting of atoms 
restricted to a single spin state.
In the strongly correlated regime, the physics controlling the possible novel phases of the condensate remains
largely unexplored, and few-body aspects can play a central role in the properties and dynamics of the 
system through manifestations of Efimov physics.
The present study solves the three-body problem for bosonic spinors using the hyperspherical adiabatic representation and characterizes the multiple 
families of Efimov states in spinor systems as well as their signatures in the scattering observables relevant for spinor 
condensates. These solutions exhibit a rich array of possible phenomena originating in universal few-body physics, which can strongly 
affect the spin dynamics and three-body mean-field contributions for spinor condensates. The collisional aspects of atom-dimer spinor 
condensates are also analyzed and effects are predicted that derive from Efimov physics.
\end{abstract}


\maketitle

In recent years, the development of optical traps has stimulated the realization of spinor condensates \cite{SpinorReview}.  
The coupling of the degenerate spin degrees of freedom leads to novel quantum phenomena such as the formation of 
spin domains, spin textures, spin mixing dynamics, and counterintuitive quantum phases. These have been intensively investigated both experimentally
\cite{SpinorExp1,SpinorExp2,SpinorExp3,SpinorExp4,SpinorExp5,SpinorExp6,SpinorExp7,SpinorExp8,SpinorExp9,SpinorExp10,SpinorExp11,SpinorExp12} 
and theoretically \cite{SpinorTheo1,SpinorTheo2,SpinorTheo3,SpinorTheo4,SpinorTheo5,SpinorTheo6,SpinorTheo7,SpinorTheo8,SpinorTheo9,SpinorTheo10}. 
Such phenomena have been shown to be sensitive to the (typically weak) interatomic interactions, characterized by multiple scattering 
lengths associated with the various atomic hyperfine spins states. 
Of particular interest is the fact that strongly correlated spinor condensates can enable explorations of spinor physics in exotic dynamical regimes. 
Although the scattering lengths for most alkali species are modest or even small, one key exception is $^{85}$Rb \cite{klausen2001PRA}. There have been several proposals to create strongly correlated spinor condensates \cite{gerbier2006PRA,zhang2009PRL,hamley2009PRA,kaufman2009PRA,tscherbul2010PRA,hanna2010NJP,
papoular2010PRA} where the scattering lengths substantially exceed the range of interatomic interactions, i.e. the van der Waals length 
$r_{\rm vdW}$.  This enables the spin states to effectively interact even at large distances. 
In this scenario one should also consider few-body correlations, notably effects associated with the existence of Efimov states 
\cite{efimov1970SJNP,braaten2006PR,wang2013AAMOP}. 

In  {\em single}-spin condensates, when the interactions are enhanced by the presence of a Feshbach resonance 
\cite{chin2010RMP}, an infinity of Efimov states emerges that strongly affects scattering observables at ultracold energies 
\cite{efimov1970SJNP,braaten2006PR,wang2013AAMOP}. More recently, advances have been made in our understanding of universal Efimov physics 
in an ultracold quantum 
gas. Despite the complex nature of the interatomic interactions, 
recent experimental \cite{IBK_Exps,LENS_Exps,Rice_Exps,Kayk_Exps,Jochim_Exps,Ohara_Exps,Ueda_Exps,JILA_Exps} and theoretical 
\cite{Chin3BP,Ueda_Theo,JILA_Theo,Schimdt,Jensen} studies have shown surprisingly that 
the usual three-body parameter is universal, depending only on $r_{\rm vdW}$, which now permits even more quantitative predictions of 
interesting few-body phenomena to be made. 

The present exploration of Efimov physics in a spinor condensate system shows that the additional spin degrees of freedom can
fundamentally modify the Efimov trimer's energy spectrum and that scattering processes can strongly affect the condensate spin dynamics. 
In the context of nuclear physics, where isospin symmetry plays an important role, the work of Bulgac and Efimov \cite{bulgac1976SJNP} 
demonstrated a much richer structure for Efimov physics when the spin degree of freedom is considered. In this case, multiple families 
of Efimov states can coexist, depending on the particular spin states and different scattering lengths in the problem
\cite{DIncao}. This is in striking contrast to the {\em standard} Efimov scenario where only a single spin state is available. For the multilevel bosonic systems examined here, 
with the topologically distinct case of a spinor condensate,
the atomic hyperfine states provide the internal atomic structure.  Several interesting effects are predicted for the three-body 
scattering observables controlling the dynamical evolution of spinor condensates. Similarly, our results point to the possibility of exploring
spinor physics in an atom-dimer mixture. These results emerge from a calculation of the collisional properties of this system, including a characterization of the signatures of Efimov physics.

The study of few-body physics in spinor condensates requires proper inclusion of the multichannel nature of interatomic interactions, 
originating from the underlying atomic hyperfine structure. Our study begins from the multichannel generalization of the zero-range Fermi pseudopotential 
\cite{mehta2008PRA,rittenhouse2010PRA,KartavsevMacek} for $s$-wave interactions (in a.u.), namely:
\begin{equation}
\hat{v}(r)=\frac{4\pi\hat{A}}{m}\delta^3(\vec{r})\frac{\partial}{\partial r}r,\label{TwoBodyPotential}
\end{equation}  
where $\delta^3(\vec{r})$ is the usual three-dimensional Dirac-$\delta$ function and $\hat{A}$ the scattering length 
{\em matrix} written in the two-body spin basis denoted by $\{|\sigma\rangle\}$.  Within this framework, the three-body 
problem is solved in the adiabatic hyperspherical representation, using the Green's function method developed in Refs. \cite{mehta2008PRA,rittenhouse2010PRA}. 
In this representation, the hyperradius $R$ determines the overall size of the system, and the internal motion is described by a set of five hyperangles, 
collectively denoted by $\Omega$. Briefly, the adiabatic fixed-$R$ eigenvalue equation reads:  
\begin{equation}
\Big[\frac{\hat\Lambda^2(\Omega)+\frac{15}{4}}{2\mu R^2}+\hat{V}(R,\Omega)+\hat E_{\Sigma}\Big]{\Phi}(R;\Omega)=U(R){\Phi}(R;\Omega),\label{Had}
\end{equation}
where $\mu=m/\sqrt{3}$ is the three-body reduced mass, $\hat\Lambda$ is the grand angular momentum operator \cite{avery}, $\hat{V}$ is the sum of 
pairwise interactions, and $\hat E_{\Sigma}$ is the sum of the atomic energy levels, which is diagonal in the three-body spin basis $\{|\Sigma\rangle\}$. The 
channel functions $\Phi(R;\Omega)$ and the three-body potentials $U(R)$ describe the physical properties of the system and are obtained by solving 
Eq.~(\ref{Had}) for fixed values of $R$. Application of the zero-range potential model reduces the problem to solving a transcendental equation 
whose roots, $s_{\nu}(R)$, determine $U_{\nu}(R)$ through
\begin{equation}
U_\nu(R)=\frac{s_\nu(R)^2-1/4}{2\mu R^2}.\label{3bp}
\end{equation}
(See outline of our formulation in Ref.~\cite{SupMat}.) For three-identical bosons,
for instance, solving Eq.~(\ref{Had}) in the limit $R/a\rightarrow 0$ yields a {\em single} imaginary root,
independent of $R$, with numerical value $s_{0}\approx1.0062i$. 
Insertion of $s_0$ into Eq. \eqref{3bp} produces the {\it{attractive}} $1/R^2$ 
potential that supports an infinity of three body bound states 
characteristic of the Efimov effect. In the present study, the threshold energy levels, $\hat E_{\Sigma}$ in Eq.({\ref{Had}) are degenerate and are set equal to zero. In 
spinor condensates at vanishingly small magnetic fields, the atomic levels are $(2f+1)$-fold degenerate ($f$ is the atomic hyperfine angular 
momentum and $m_{f}=-f,...,f,$ its azimutal component).  In fact, this degeneracy leads to fundamentally different three-body physics 
than is obtained for the usual Efimov case with atoms in a single spin state.

The interatomic interaction for spinor condensates \cite{SpinorReview} is spin-dependent, and we assume the scattering length operator 
in Eq.~(\ref{TwoBodyPotential}) can be represented as
\begin{eqnarray}
\hat{A}=\sum_{F_{\rm 2b}M_{F_{\rm 2b}}} 
a_{F_{\rm 2b}} 
|F_{\rm 2b}M_{F_{\rm 2b}} \rangle 
\langle F_{\rm 2b}M_{F_{\rm 2b}} |,\label{A2b}
\end{eqnarray} 
where $F_{\rm 2b}$ and $M_{F_{\rm 2b}}$ are the two-body total spin and its projection. Due to bosonic symmetry only the symmetric spin states 
($F_{\rm 2b}$$\equiv$even) are allowed to interact with rotationally-invariant scattering lengths $\{a_{0},a_{2},...,a_{2f}\}$.  
These scattering lengths set important length scales in the problem, and their relative magnitudes and signs determine 
many-body properties such as the miscibility of the different spin components. Moreover, the scattering lengths also determine
the nature of the three-body interactions and many of the scattering properties of the system, potentially impacting the spin dynamics of 
condensates.
\begin{figure}[htbp]
\includegraphics[width=3.0in]{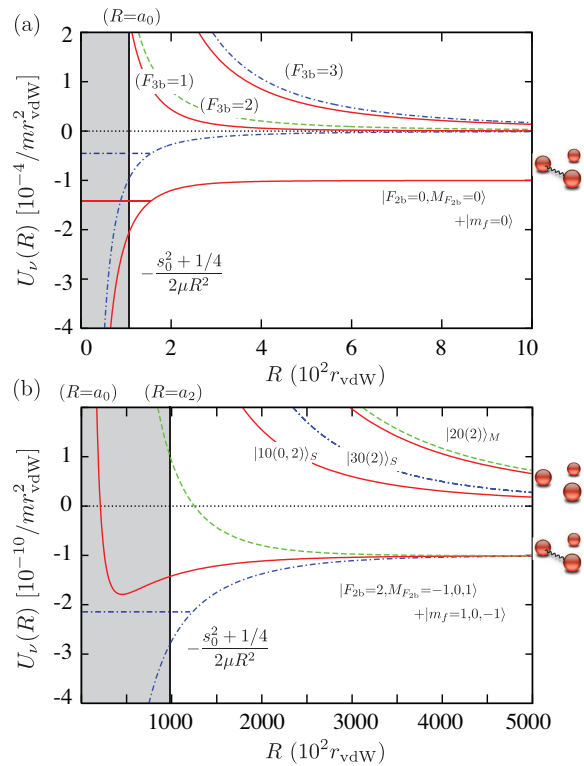}
\caption{(Color online) $F_{\rm 3b}=$1 (red solid line), 2 (green dashed line), and 3 (blue dash-dotted line) hyperspherical adiabatic potentials for $f$$=$$1$ 
spinors with $a_{0}$$=$$10^2 r_{\rm vdW}$ and $a_{2}$$=$$10^5r_{\rm vdW}$. 
(a) For $R$$\le$$\{a_0,a_2\}$ (shaded region) two attractive potentials exist (both with $s_0$$\approx$$1.0062i$), allowing for 
two families of Efimov states, and for $R$$>$$a_{0}$, one of these potentials turns into an atom-dimer channel, 
$|F_{\rm 2b}$$=$$0,M_{F_{\rm 2b}}$$=$$0\rangle$+$|m_{f}$$=$$0\rangle$. (b) For $a_{0}$$\le$$R$$\le$$a_2$ (shaded region) only one family of Efimov states exist
($s_0$$\approx$$1.0062i$) and for $R$$\gg$$a_{2}$ three (asymptotically degenerate) potentials describe atom-dimer channels,
$|F_{\rm 2b}$$=$$2,M_{F_{\rm 2b}}$$=$$-1,0,1\rangle$+$|m_{f}$$=$$1,0,-1\rangle$.}\label{Pot_f=1}
\end{figure}

Figure~\ref{Pot_f=1} shows the three-body potentials for $f$$=$$1$ atoms for the allowed values of the total three-body hyperfine spin, 
$|$$F_{\rm 2b}$$-$$f$$|$$\le$$F_{\rm 3b}$$\le$$F_{\rm2b}$$+$$f$.
These are, of course, independent of $M_{\rm 3b}=M_{F_{\rm 2b}}+m_{f}$. [The $F_{\rm 3b}$$=$$0$ states are spatially antisymmetric and thus noninteracting in
the potential model 
of Eq.~(\ref{TwoBodyPotential}).] The results in Fig.~\ref{Pot_f=1} were obtained by solving Eq.~(\ref{Had}) in the spin basis 
$\{|\Sigma\rangle\}$ \cite{SupMat} and with $a_{0}$$=$$10^2 r_{\rm vdW}$ and $a_{2}$$=$$10^5r_{\rm vdW}$. 
Figure~\ref{Pot_f=1}(a) emphasizes the three-body physics for $R$$\le$$\{a_0,a_2\}$ (shaded region) where {\em two} attractive potentials 
exist for $F_{\rm 3b}=1$ (red solid line) and 3 (blue dash-dotted line). Both potentials are associated with $s_0$$\approx$$1.0062i$ and allow 
for the coexistence of {\em two} families of Efimov states, 
(represented in Fig.~\ref{Pot_f=1} by the horizontal solid and dashed-dotted lines) 
---a feature absent in systems of single state atoms. 
For $R$$>$$a_{0}$, the
$F_{\rm 3b}$$=$$1$ potential turns into an atom-dimer channel describing collisions between a $|F_{\rm 2b}$$=$$0,M_{F_{\rm 2b}}$$=$$0\rangle$ 
dimer, with energy $-1/ma_0^2$, and a $|m_{f}$$=$$0\rangle$ atom. For $a_{0}$$\le$$R$$\le$$a_2$ [shaded region in Fig.~\ref{Pot_f=1}(b)] only 
the $F_{\rm 3b}$$=$$3$ family of Efimov states persists. For $R$$\gg$$a_{2}$, this $F_{\rm 3b}$$=$$3$ potential, and two other $F_{\rm 3b}$$=$$1$ 
and $F_{\rm 3b}$$=$$2$ potentials,
converge to the dimer energy $-1/ma_2^2$ and describe atom-dimer collisions in states $|F_{\rm 2b}$$=$$2,M_{F_{\rm 2b}}$$=$$-1,0,1\rangle$+$|m_{f}$$=$$1,0,-1\rangle$. 
This offers an interesting scenario in which one can study atom-dimer 
spinor mixtures, some of whose collisional properties are described in Ref.~\cite{SupMat}. 
In Fig.~\ref{Pot_f=1}(b), the repulsive potentials for $R$$\gg$$a_2$ describe collisions 
between three free atoms in the symmetric spin states $|F_{\rm 3b},M_{F_{\rm 3b}}(F_{\rm 2b})\rangle=|30(2)\rangle_{S}$ and $|10(0,2)\rangle_{S}$ and the mixed symmetry 
state $|20(2)\rangle_{M}$ (see Ref.~\cite{SupMat}). Note that only $F_{\rm 3b}$$=$1 states are sensitive to both $a_0$ and $a_2$.
\begin{table}[htbp]
\caption{Values of $s_{\nu}$ relevant for $f$$=$$1$ and $2$ spinor condensates covering all possible regions of $R$  for the different ranges 
of the relevant scattering lengths. For $f$$=$1 we list the lowest few values of $s_{\nu}$ for each $F_{\rm 3b}$ 
while for $f$=2 we only list the values of $s_{\nu}$ and their multiplicity (superscript), instead of the specific value of $F_{\rm 3b}$ where they occur. 
}\label{TabRoots}
\begin{ruledtabular}
\begin{tabular}{lccc}
  {($f=1$)}  & $F_{\rm 3b}=1$ &$ F_{\rm 3b}=2$ & $F_{\rm 3b}=3$ \\
[0.05in]\hline
$R$$\ll$$|a_{\{0,2\}}|$  & $1.0062i,2.1662$ & $2.1662$ & $1.0062i,4.4653$ \\
$|a_0|$$\ll$$R$$\ll$$|a_2|$   & $0.7429$  & $2.1662$ & $1.0062i,4.4653$ \\
 $|a_2|$$\ll$$R$$\ll$$|a_0|$  & $0.4097$  &  $4$ & $2$ \\
$R$$\gg$$|a_{\{0,2\}}|$  & $2$  &  $4$  & $2$ \\
 [0.05in]\hline
  {($f=2$)}  & \multicolumn{3}{c}{$F_{\rm 3b}=0,1,...,6$}\\
[0.05in]\hline
$R$$\ll$$|a_{\{0,2,4\}}|$  & \multicolumn{3}{c}{$1.0062i^{(5)},2.1662^{(5)}$} \\ 
$|a_0|$$\ll$$R$$\ll$$|a_{\{2,4\}}|$ &  \multicolumn{3}{c}{$1.0062i^{(4)},0.4905^{(1)}$}   \\ 
$|a_2|$$\ll$$R$$\ll$$|a_{\{0,4\}}|$  & \multicolumn{3}{c}{$1.0062i^{(1)},0.7473i^{(1)},0.6608^{(1)}$} \\
$|a_4|$$\ll$$R$$\ll$$|a_{\{0,2\}}|$  & \multicolumn{3}{c}{$1.0062i^{(1)},0.5528i^{(1)},0.3788i^{(1)},0.5219^{(1)}$}   \\
$|a_{\{0,2\}}|$$\ll$$R$$\ll$$|a_4|$ &  \multicolumn{3}{c}{$1.0062i^{(1)},0.6608^{(1)}$} \\
$|a_{\{0,4\}}|$$\ll$$R$$\ll$$|a_2|$ & \multicolumn{3}{c}{$1.0062i^{(1)},0.5528i^{(1)},0.5219^{(1)}$} \\
$|a_{\{2,4\}}|$$\ll$$R$$\ll$$|a_0|$ & \multicolumn{3}{c}{$0.6861^{(1)}$}\\
$R$$\gg$$|a_{\{0,2,4\}}|$  & \multicolumn{3}{c}{$2^{(5)},4^{(2)}$} \\ 
\end{tabular}
\end{ruledtabular}
\end{table}

Table \ref{TabRoots} summarizes the values of $s_{\nu}$ relevant for $f$$=$$1$ and $2$ spinor condensates, covering all possible regions of $R$ 
and for different magnitudes of the relevant scattering lengths. For $f$$=$$1$ the values of $s_{\nu}$ are listed according to the value of $F_{\rm 3b}$
while for $f$$=$$2$ they are not assigned in detail (see the complete assignment in Ref.~\cite{SupMat}).  
Notably, for $f$$=$$1$ ensembles, the imaginary values of $s_{\nu}$ agree exactly with the ones for single level atoms, 
except with the important distinction that such roots can be degenerate in the spinor case. 
It is well known that the existence of overlapping series of states can lead to formation of ultra long-lived states \cite{wang1991PRA}. 
In our present case, the $F_{\rm 3b}=1$ and 3 Efimov states can interact for finite (but small) magnetic fields and such controllability can 
not only produce long-lived states but also, due to their weakly bound character, affect the spin dynamics of the condensate.
The occurrence of such effects, however, will depend on the three-body short-range physics \cite{JILA_Theo}. Further analysis of this 
parameter space is beyond the scope of the present study.  This feature opens up exciting possibilities for the study of Efimov trimers in spinor 
condensates in a well-controlled manner. More interestingly, $f$$=$$2$ ensembles exhibit several values of $s_{\nu}$ that differ from 
those for single level atoms. The physics controlling the appearance of these new roots is due to the fact that the atoms 
in the two-body states $|F_{\rm 2b}M_{F_{\rm 2b}}\rangle$ are not in a pure quantum state. Instead, they are in a mixture of states, with the amount of mixing 
controlled by the angular momentum algebra.

\begin{table*}[htbp]
\caption{Scattering length dependence for $F_{\rm 3b}=1$, 2 and 3 three-body scattering observables relevant for $f$=$1$ spinor condensates. 
Here, $K_{3}$ is the three-body recombination rate [superscripts $(0)$ and $(2)$ indicating recombination into weakly bound $F_{\rm 2b}=0$ and 2 
dimers and $(d)$ recombination rate into deeply bound states] and $a_{\rm 3b}$ the three-body scattering length matrix element. $M(a)$, $P(a)$ and 
$T(a)$ are given in Eqs.~(\ref{Mformula})-(\ref{Tformula}). Here, $\gamma$ is a universal constant that can be calculated for each observable, 
$s_1\approx0.7429$ ($|a_2|$$\gg$$|a_0|$) and $s_1\approx0.4097$ ($|a_0|$$\gg$$|a_2|$) 
for $F_{\rm 3b}$$=$$1$ and for $F_{\rm 3b}$$=$2, $s_1\approx2.1662$ and $k^2=2\mu E$.}\label{TabScat}
\begin{ruledtabular}
\begin{tabular}{cccccc|ccc}
$(F_{\rm 3b}=1)$ & $a_2\gg a_0$ & $a_2\gg |a_0|$ & $|a_2|\gg a_0$ & $|a_2|\gg |a_0|$ &~& $(F_{\rm 3b}=2)$ & $a_2\gg r_{\rm vdW}$ & $|a_2|\gg r_{\rm vdW}$ \\
[0.05in] \hline 
$K_{3}^{(0)}/a_{2}^4$       & {\scriptsize$M_{s_0}^{\eta}(a_0) \left(\frac{a_0}{a_2}\right)^{2s_1}$}& --- & {\scriptsize$M_{s_0}^{\eta}(a_0) \left(\frac{a_0}{a_2}\right)^{2s_1}$} & --- && $K_{3}^{(2)}/a_{2}^8$        & {\scriptsize$\gamma k^4$} & --- \\
$K_{3}^{(2)}/a_{2}^4$       & {\scriptsize$\gamma$$+$$M_{s_0}^\eta(a_0) \left(\frac{a_0}{a_2}\right)^{4s_1}$} & {\scriptsize$\gamma$$+$$P_{s_0}^\eta(a_0)\left(\frac{a_0}{a_2}\right)^{4s_1}$} & --- & --- && $K_{3}^{(d)}/a_{2}^8$        &{\scriptsize$\gamma k^4$} & {\scriptsize$\gamma k^4\left(\frac{r_{\rm vdW}}{a_2}\right)^{2s_1}$}\\
$K_{3}^{(d)}/a_{2}^4$       & {\scriptsize$\gamma\left(\frac{a_0}{a_2}\right)^{2s_1}$}  &{\scriptsize$P_{s_0}^\eta(a_0)\left(\frac{a_0}{a_2}\right)^{2s_1}$} &  {\scriptsize$\gamma\left(\frac{a_0}{a_2}\right)^{2s_1}$} & {\scriptsize$P_{s_0}^\eta(a_0)\left(\frac{a_0}{a_2}\right)^{2s_1}$} && $a_{\rm 3b}^{(2)}/a_{2}^4$  & --- & ---\\
$a^{(1)}_{\rm 3b}/a_{2}^4$ & {\scriptsize$\gamma$$+$$O_{s_0}^{\eta}(a_0)\left(\frac{a_0}{a_2}\right)^{4s_1}$} & {\scriptsize$\gamma$$+$$T_{s_0}^{\eta}(a_0)\left(\frac{a_0}{a_2}\right)^{4s_1}$}& {\scriptsize$\gamma$$+$$O_{s_0}^{\eta}(a_0)\left(\frac{a_0}{a_2}\right)^{4s_1}$} & {\scriptsize$\gamma$$+$$T_{s_0}^{\eta}(a_0)\left(\frac{a_0}{a_2}\right)^{4s_1}$} \\
[0.05in] \hline 
$(F_{\rm 3b}=1)$ & $a_0\gg a_2$ & $a_0\gg |a_2|$ & $|a_0|\gg a_2$ & $|a_0|\gg |a_2|$ && $(F_{\rm 3b}=3)$ & $a_2\gg r_{\rm vdW}$ & $|a_2|\gg r_{\rm vdW}$\\
[0.05in] \hline 
$K_{3}^{(0)}/a_{0}^4$       & {\scriptsize$\gamma$$+$$M_{s_0}^\eta(a_2) \left(\frac{a_2}{a_0}\right)^{4s_1}$} & {\scriptsize$\gamma$$+$$P_{s_0}^\eta(a_2)\left(\frac{a_2}{a_0}\right)^{4s_1}$} & --- & --- && $K_{3}^{(2)}/a_{2}^4$ &  {\scriptsize$M_{s_0}^{\eta}(a_2)$} & ---\\
$K_{3}^{(2)}/a_{0}^4$       & {\scriptsize$M_{s_0}^{\eta}(a_2) \left(\frac{a_2}{a_0}\right)^{2s_1}$}& --- & {\scriptsize$M_{s_0}^{\eta}(a_2) \left(\frac{a_2}{a_0}\right)^{2s_1}$} & --- && $K_{3}^{(d)}/a_{2}^4$ & $\gamma$ & {\scriptsize$P_{s_0}^{\eta}(a_2)$} \\
$K_{3}^{(d)}/a_{0}^4$       & {\scriptsize$\gamma\left(\frac{a_2}{a_0}\right)^{2s_1}$}  &{\scriptsize$P_{s_0}^\eta(a_2)\left(\frac{a_2}{a_0}\right)^{2s_1}$} &  {\scriptsize$\gamma\left(\frac{a_2}{a_0}\right)^{2s_1}$} & {\scriptsize$P_{s_0}^\eta(a_2)\left(\frac{a_2}{a_0}\right)^{2s_1}$} && $a_{\rm 3b}^{(3)}/a_{2}^4$ & {\scriptsize$O_{s_0}^{\eta}(a_2)$} & {\scriptsize$T_{s_0}^{\eta}(a_2)$}\\
$a^{(1)}_{\rm 3b}/a_{0}^4$ & {\scriptsize$\gamma$$+$$O_{s_0}^{\eta}(a_2)\left(\frac{a_2}{a_0}\right)^{4s_1}$} & {\scriptsize$\gamma$$+$$T_{s_0}^{\eta}(a_2)\left(\frac{a_2}{a_0}\right)^{4s_1}$}& {\scriptsize$\gamma$$+$$O_{s_0}^{\eta}(a_2)\left(\frac{a_2}{a_0}\right)^{4s_1}$} & {\scriptsize$\gamma$$+$$T_{s_0}^{\eta}(a_2)\left(\frac{a_2}{a_0}\right)^{4s_1}$} 
\end{tabular}
\end{ruledtabular}
\end{table*}

The above results illustrate the rich structure of Efimov states in spinor systems. Such richness also appears
in the three-body scattering observables. Here, we use a WKB model \cite{DIncao} to determine the scattering length 
and energy dependence of collision rates for all relevant scattering processes for $f$$=$$1$ spinor condensates. 
The results for $F_{\rm 3b}=$ 1, 2 and 3, are summarized in Table \ref{TabScat}.
Notice that scattering observables can display log-periodic interference and resonant effects due to the multiple families 
of Efimov states and collision pathways available in spinor ensembles. Interference and resonance effects 
are parameterized according to, respectively,
\begin{align}
&M_{s_0}^{\eta}(a)=\alpha~{e^{-2\eta}}\left[\sin^2\left(\mbox{$|s_{0}|\ln \frac{a}{r_{\phi}}$}\right)+\sinh^2\eta\right],\label{Mformula}\\ 
&P_{s_0}^{\eta}(a)=\beta~\frac{\sinh2\eta}{\sin^2\left(|s_{0}|\ln \frac{a}{r_{\phi}}\right)+\sinh^2\eta}\label{Pformula},
\end{align} 
where $r_{\phi}$$=$$r_{\rm vdW}e^{-\phi/|s_0|}$ is the three-body parameter, incorporating the short-range physics through the phase $\phi$ \cite{JILA_Theo}, 
and $\eta$ is the three-body inelasticity parameter \cite{braaten2006PR,wang2013AAMOP} which encapsulates the probability for decay into deeply bound molecular
states. 
In the above equations, $\alpha$ and $\beta$ are universal constants that can be evaluated for each $F_{\rm 3b}$.
In Table \ref{TabScat}, $K_{3}^{(0)}$ and $K_{3}^{(2)}$ denote the collision rate for three-body recombination into weakly bound $F_{\rm 2b}=0$ and $2$ dimers, 
respectively. Such dimers can still remain trapped and further dissociate into free atoms via collision with other atoms with dissociation rates 
$D_{3}^{(0)}\propto K^{(0)}_{3}(k^4a_{0})$ and $D_{3}^{(2)}\propto K^{(2)}_{3}(k^4a_2)$, where $k^2=2\mu E$ with $E>0$ being the three-body collision energy. 
This interplay between dimer formation and dissociation can provide an interesting dynamical regime in spinor condensates which is 
absent when the scattering lengths are small. Three-body recombination into deeply bound molecular states, $K_{3}^{(d)}$, 
can only lead to losses and can display resonant enhancements due to formation of Efimov states or can be suppressed due to repulsive 
three-body interactions. 
Note that the total rate is obtained by multipling $K_{3}$ in Table \ref{TabScat} by the appropriate factors that account 
for the various degeneracies in the problem.

While inelastic collisions determine the stability of condensates, three-body {\em elastic} processes determine how 
different spins interact. Moreover, they can also display resonant effects due to Efimov states, which can strongly affect the spin dynamics. 
Within the mean-field description of spinor condensates \cite{SpinorReview}, such effects can be incorporated through
the {\em three-body} scattering length operator 
\begin{align}
\hat{A}_{\rm 3b}=
\hspace{-0.1in}
\sum_{F_{\rm 3b}M_{F_{\rm 3b}}\atop F_{\rm 2b}}
\hspace{-0.1in}
a_{\rm 3b}^{(F_{\rm 3b})}|F_{\rm 3b}M_{F_{\rm 3b}}(F_{\rm 2b})\rangle \langle F_{\rm 3b}M_{F_{\rm 3b}}(F_{\rm 2b})|.\label{A3b}
\end{align}
This is a natural extension of the three-body scattering length (units of length$^4$) as defined in \cite{efimov1970SJNP,BraatenNieto1999,BraatenHammerMehen2002,
BedaqueBulgacRupak2003,bulgac2002PRL}. The scattering length dependence of $a_{\rm 3b}$ is listed in Table \ref{TabScat} for $f$$=$$1$ atoms, showing the 
different ways it is influenced by Efimov physics. (Note that for $F_{\rm 3b}$=2, $a_{\rm 3b}$ is not defined since for this case a higher centrifugal barrier
suppresses collisions at ultracold energies \cite{SupMat}.)
One interesting case emerges, for instance, when $a_0$$<$$0$ or $a_2$$<$$0$ where $a_{\rm 3b}$ can display resonant effects if $a_{0}$ or $a_2$ 
are tuned near a Efimov resonance ---parametrized in Table \ref{TabScat} by the tangent-like, log-periodic, function
\begin{align}
&T_{s_0}^{\eta}(a)=\alpha+\beta~\frac{2\sin\left(|s_{0}|\ln \frac{a}{r_{\phi}}\right)\cos\left(|s_{0}|\ln \frac{a}{r_{\phi}}\right)}{\sin^2\left(|s_{0}|\ln\frac{a}{r_{\phi}}\right)+\sinh^2\eta},\label{Tformula}
\end{align}
where $\alpha$ and $\beta$ are, again, universal constants. [Note that oscillations in $a_{\rm 3b}$, parametrized by $O^{\eta}_{s_{0}}(a)$ in Ref.~\cite{SupMat},
are also allowed.] In this case $|a_{\rm 3b}|^\frac{1}{4}\gg\{|a_0|,|a_2|\}$ and three-body correlations can dominate mean-field interactions, allowing for both
attractive ($a_{\rm 3b}$$<$$0$) and repulsive ($a_{\rm 3b}$$>$$0$) three-body interactions. 

From the mean-field perspective, in order to understand three-body contributions for spinor condensates, it is convenient to
write $\hat A_{\rm 3b}$ in a way that makes explicit the importance of spin-exchange interactions \cite{SpinorTheo1}. For $f$=1 atoms 
we can rewrite Eq.~(\ref{A3b}) as \cite{SupMat}
\begin{eqnarray}
\hat A_{\rm 3b}=\alpha_{\rm 3b}+\alpha_{\rm 3b}^{\rm ex}\sum_{i<j}\vec{f}_{i}\cdot\vec{f}_{j},\label{A3bSpinor}
\end{eqnarray}
where $\vec{f}_i$ ($i$=1, 2 and 3) is the atomic hyperfine angular momentum for the atom 
$i$, and the three-body direct and exchange interactions given, respectively, by
\begin{eqnarray}
\alpha_{\rm 3b}=({2a_{\rm 3b}^{(3)}+3a_{\rm 3b}^{(1)}})/{5}\\
\alpha_{\rm 3b}^{\rm ex}=({a_{\rm 3b}^{(3)}-a_{\rm 3b}^{(1)}})/{5}.
\end{eqnarray}
This form for $\hat A_{\rm 3b}$ is in close analogy to the two-body spinor case, in which 
$\hat A_{\rm 2b}$$=$$\alpha_{\rm 2b}$$+$$\alpha_{\rm 2b}^{\rm ex}\vec{f}_{1}\cdot\vec{f}_{2}$, 
where $\alpha_{\rm 2b}$$=$$($$a_0$$+$$3a_2$$)$$/$$3$ and $\alpha_{\rm 2b}^{\rm ex}$$=$$($$a_2$$-$$a_0$$)$$/$$3$ \cite{SpinorTheo1}. 
Mean-field contributions, however, are introduced through the corresponding two- and three-body coupling constants $g_{\rm 2b}$$=$$(4\pi/m)\alpha_{\rm 2b}$,
$g_{\rm 2b}^{\rm ex}$$=$$(4\pi/m)\alpha_{\rm 2b}^{\rm ex}$, $g_{\rm 3b}$$=$$3^{\frac{1}{2}}(12\pi/m)\alpha_{\rm 3b}$, and
$g_{\rm 3b}^{\rm ex}$$=$$3^{\frac{1}{2}}(12\pi/m)\alpha_{\rm 3b}^{\rm ex}$, respectively \cite{bulgac2002PRL}.

Observe that the $a_0^4$ (or $a_2^4$) dependence of $g_{\rm 3b}$ and $g_{\rm 3b}^{\rm ex}$ can quickly make the three-body direct and spin-exchange mean-field 
energies, $g_{\rm 3b}n^2$ and $g^{\rm ex}_{\rm 3b}n^2$, comparable to their two-body counterparts, $g_{\rm 2b}n$ and $g^{\rm ex}_{\rm 2b}n$.
In fact, resonant effects in $\alpha_{\rm 3b}^{\rm ex}$ due to Efimov states can strongly affect both ferromagnetic ($g_{\rm 2b}^{\rm ex}$$<$$0$) and 
antiferromagnetic ($g_{\rm 2b}^{\rm ex}$$>$$0$) phases in spinor condensates \cite{SpinorReview,SpinorTheo1} whenever $|g^{\rm ex}_{\rm 3b}|n^2>|g^{\rm ex}_{\rm 2b}|n$ 
with $g_{\rm 3b}^{\rm ex}$ and $g_{\rm 2b}^{\rm ex}$ having {\em opposite} signs. 
In particular, in the ferromagnetic phase, we speculate that attractive two-body interactions, $g_{\rm 2b}$$<$$0$ and $g_{\rm 2b}^{\rm ex}$$<$$0$, can be stabilized by 
a repulsive three-body interaction ($g_{\rm 3b}$$>$$0$ and $g_{\rm 3b}^{\rm ex}$$>$$0$) to form local, self-bound, quantum droplets of spinor characters
in a similar spirit of Ref.~\cite{bulgac2002PRL}. 
The study of mean-field three-body contributions and their possible effects in spinor condensates, however, will be a subject of future investigations.

Finally, our present study has also shown the possibility of creating atom-dimer spinor condensates where $F_{\rm 2b}\ne0$ dimers 
can exchange $M_{F_{\rm 2b}}$ by colliding with other atoms in $|m_{f}\rangle$ states. For instance, for $f$$=$$1$ atoms, 
Fig.~\ref{Pot_f=1}(b) shows $F_{\rm 2b}=2$ dimers can collide with atoms in states $|$$-$$1\rangle$, $|0\rangle$ and $|1\rangle$ and their collisional
properties are listed in Table SI of Ref. \cite{SupMat}. Similar to atomic spinor condensates, in the mean-field approximation the relevant 
parameters for this atom-dimer mixture is the elastic atom-dimer scattering length matrix $\hat A_{\rm ad}$ whose elements $a_{\rm ad}$ are listed
in Table~SI. As one can see, Efimov resonances can also strongly affect the mean-field energy. Moreover, if both $a_{0}>0$ and $a_2>0$ are 
large, both $F_{\rm 2b}=0$ and 2 dimers can remain trapped, leading to an interesting regime where reactive scattering can affect the dynamics of the system \cite{SupMat}.

In summary, we have explored universal aspects of Efimov physics in spinor systems and found a rich variety of scattering phenomena that strongly 
affect the spin dynamics in strongly correlated spinor condensates. The multiple, co-existing, families of Efimov states characteristic of 
spinor systems can lead to non-trivial spin dynamics, dominated by three-body correlations, as well as allowing for the existence of ultralong 
lived Efimov states. We also study few-body aspects of atom-dimer spinor condensates and show that it can offer novel regimes for studying 
spin-like physics.

This work was supported by the U. S. National Science Foundation and by an AFOSR-MURI grant.

\setcounter{equation}{0}
\setcounter{figure}{0}
\setcounter{table}{0}

\renewcommand{\theequation}{S\arabic{equation}}
\renewcommand{\thefigure}{S\arabic{figure}}
\renewcommand{\thetable}{S\Roman{table}}

\section*{SUPPLEMENTARY MATERIAL}

\section{Two- and Three-body spin functions}\label{SecI}

In this section we will give explicit expressions for both two- and three-body
spin states for $f=1$ atoms as well as discuss some symmetry properties relevant
for our present study. As usual, the two-body spin functions of total angular momentum $|f_{1}-f_2|\le F_{\rm 2b} \le f_1+f_2$
and projection $M_{F_{\rm 2b}}=m_{f_1}+m_{f_2}$ can be expressed in terms
of the Clebsch-Gordan coefficients as
\begin{align}
&|F_{\rm 2b}M_{F_{\rm 2b}} \rangle \equiv |(f_1f_2)F_{\rm 2b}M_{F_{\rm 2b}} \rangle=\nonumber\\
&\sum_{m_{f_{1}}m_{f_{2}}}\langle f_{1}m_{f_{1}}f_{2}m_{f_{2}}|F_{\rm 2b}M_{F_{\rm 2b}}\rangle|m_{f_{1}},m_{f_{2}}\rangle.\label{2Spin}
\end{align} 
For $f=1$ atoms these spin functions are, for $F_{\rm 2b}=0$, 1 and 2,
\begin{align}
&\mbox{$|00\rangle=\frac{|-1,1\rangle}{\sqrt{3}}-\frac{|0,0\rangle}{\sqrt{3}}+\frac{|1,-1\rangle}{\sqrt{3}}$}, \\
&\mbox{$|10\rangle=-\frac{|-1,1\rangle}{\sqrt{2}}+\frac{|1,-1\rangle}{\sqrt{2}}$}, \\
&\mbox{$|1$$\pm$$1\rangle=\mp\frac{|0,\pm1\rangle}{\sqrt{2}}\pm\frac{|\pm1,0\rangle}{\sqrt{2}}$}, \\
&\mbox{$|20\rangle=\frac{|-1,1\rangle}{\sqrt{6}}+\frac{|0,0\rangle}{\sqrt{3/2}}+\frac{|1,-1\rangle}{\sqrt{6}}$}, \\
&\mbox{$|2$$\pm$$1\rangle=\frac{|0,\pm1\rangle}{\sqrt{2}}+\frac{|\pm1,0\rangle}{\sqrt{2}}$}, \\
&\mbox{$|2$$\pm$$2\rangle=|$$\pm$$2,$$\pm$$2\rangle$}.
\end{align}
As one can see, $F_{\rm 2b}=1$ states are antisymmetric and must be excluded from our $s$-wave interaction model.

Similarly, three-body spin functions of total angular momentum $|F_{\rm 2b}-f_3|\le F_{\rm 3b} \le F_{\rm 2b}+f_3$
and projection $M_{F_{\rm 3b}}=M_{F_{\rm 2b}}+m_{f_3}$ can be expressed in terms of the corresponding two spin states Eq. (\ref{2Spin})
as
\begin{align}
&|F_{\rm 3b}M_{F_{\rm 3b}}(F_{\rm 2b})\rangle\equiv |(f_1f_2f_3)F_{\rm 3b}M_{F_{\rm 3b}}(F_{\rm 2b})\rangle=\nonumber\\
&\sum_{M_{F_{\rm 2b}},m_{f_3}}
\langle F_{\rm 2b}M_{F_{\rm 2b}}f_3 m_{f_3}|F_{\rm 3b}M_{F_{\rm 3b}}\rangle
|F_{\rm 2b}M_{F_{\rm 2b}}\rangle|m_{f_3}\rangle.
\end{align}
We will, however, analyse the three-body spin functions for $f=1$ atoms for each value of $F_{\rm 3b}$ separately since
symmetry considerations that allow us to disregard some states are less evident than in the two-body case.
Here we will consider only states with $M_{F_{\rm 3b}}=0$ but the same considerations also applies for $M_{F_{\rm 3b}}\ne0$.

For $F_{\rm 3b}=0$ one can, by inspection, determine that the corresponding spin function
\begin{align}
&\mbox{$|00(1)\rangle=\frac{|1-1\rangle|1\rangle}{\sqrt{3}}-\frac{|10\rangle|0\rangle}{\sqrt{3}}
                       +  \frac{|11\rangle|-1\rangle}{\sqrt{3}}$},
\end{align}
is antisymmetric under permutations of any two spins. Similar to the two-body case, a three-body antisymmetric spin 
state requires an antisymmetric spacial wave function in order to form a symmetric total wave function. Since in our model only 
$s$-wave interactions are allowed, antisymmetric three-body states are noninteracting, and we neglect them from our analysis and 
calculations.

For $F_{\rm 3b}=1$, the analysis is more complicated. Now, there exist three spin states with $F_{\rm 3b}=1$, each one corresponding
to the allowed values for $F_{\rm 2b}$. They are given by,
\begin{align}
&\mbox{$|10(0)\rangle=|00\rangle|0\rangle$}, \\
&\mbox{$|10(1)\rangle=-\frac{|1-1\rangle|1\rangle}{\sqrt{2}}
                       +  \frac{|11\rangle|-1\rangle}{\sqrt{2}}$}, \\
&\mbox{$|10(2)\rangle=\frac{|2-1\rangle|1\rangle}{\sqrt{10/3}}-\frac{|20\rangle|0\rangle}{\sqrt{5/2}}
                       +  \frac{|21\rangle|-1\rangle}{\sqrt{10/3}}$}.
\end{align}
Besides the fact that permutations of spins 1 and 2 are symmetric ($F_{\rm 2b}$=even) or antisymmetric ($F_{\rm 2b}$=odd), no clear
symmetry property can be derived by inspection. The three spin functions for $F_{\rm 3b}=1$ 
can be symmetryzed to form a pair of mixed symmetry states (one symmetric and other antisymmetric with respect to permutations
of spins 1 and 2) and a fully symmetric spin state \cite{Hamermesh}, given by
\begin{align}
&\mbox{$|10(0,2)\rangle=\frac{2}{3}|10(2)\rangle+\frac{\sqrt{5}}{3}|10(0)\rangle$}.
\end{align}
Although in our calculations it is crucial include all these states, we 
determined that collision processes involving three atoms in mixed symmetry states are suppressed at low energies due to stronger centrifugal 
barriers in the three-body potentials. Therefore, in this case, the $F_{\rm 3b}=1$ totally symmetric state is dominant. For atom-dimer collisions,
however, both symmetric and mixed symmetry states should be considered. For $F_{\rm 3b}=2$ states, the corresponding spin functions
\begin{align}
&\mbox{$|20(1)\rangle=\frac{|1-1\rangle|1\rangle}{\sqrt{6}}+\frac{|10\rangle|0\rangle}{\sqrt{3/2}}
                       +  \frac{|11\rangle|-1\rangle}{\sqrt{6}}$},\\
&\mbox{$|20(2)\rangle=-\frac{|2-1\rangle|1\rangle}{\sqrt{2}}
                       +  \frac{|21\rangle|-1\rangle}{\sqrt{2}}$},
\end{align}
form a pair of mixed symmetry states. Similar to $F_{\rm 3b}=1$, we determined that for $F_{\rm 3b}=2$, collision processes involving three 
atoms in such states are also suppressed at low energies due to stronger centrifugal barriers in the three-body potentials. Consequently, they 
are neglected for the analysis of collisions between three atoms but are included in atom-dimer collisions. 
For $F_{\rm 3b}=3$, there exists only one spin function
\begin{align}
&\mbox{$|30(2)\rangle=\frac{|2-1\rangle|1\rangle}{\sqrt{5}}+\frac{|20\rangle|0\rangle}{\sqrt{5/3}}
                       +  \frac{|21\rangle|-1\rangle}{\sqrt{5}}$},
\end{align}
that forms a totally symmetric spin state and, therefore, it is included in our analysis.

\section{Three-body Green's function approach for spinor systems}

In the present study we use the Green's functions formulation developed in Refs.~\cite{xmehta2008PRA,xrittenhouse2010PRA} to solve the three-body spinor 
problem in the adiabatic hyperspherical representation. Here, we briefly outline the main steps of this formalism for spinor systems.
We solve Eq.~(2) of the main text [setting $\hat E_{\Sigma}=0$]
by expressing the corresponding Lippmann-Schwinger equation \cite{xmehta2008PRA} for each component of the channel
function $\Phi$ as
\begin{eqnarray}
\Phi_{\Sigma}(R;\Omega)=-2\mu R^2\sum_{\Sigma',k}\int d\Omega'G_{\Sigma\Sigma}(\Omega,\Omega')\nonumber\\
\times v^{(k)}_{\Sigma\Sigma'}(R,\Omega')\Phi_{\Sigma'}(R;\Omega'),\label{LS}
\end{eqnarray}
where 
\begin{eqnarray}
v^{(k)}_{\Sigma\Sigma'}(R,\Omega)=\langle\Sigma|v(r_{ij})|\Sigma'\rangle,
\end{eqnarray}
is the interaction term, given in our present study by Eq.~(1) of the main text. 
The superscript $(k)$ means that particle $k$ is a spectator while particles $i$ and $j$ interact.
The Green's functions are set to satisfy $[\hat\Lambda-(s^2-4)]G_{\Sigma\Sigma}(\Omega,\Omega')=\delta(\Omega,\Omega')$,
where $2\mu R^2 U(R)=s^2-1/4$ \cite{xmehta2008PRA}.
Our choice for the three-body spin basis is the one formed by product states 
$\{|\Sigma\rangle\}=\{|m_{f_1}m_{f_2}m_{f_3}\rangle\}$. This choice simplifies 
the formulation and, as we will see, the solutions for a given value of $F_{\rm 3b}$ and 
$M_{F_{\rm 3b}}$ can be obtained in the last step of our formulation.

As shown in Ref.~\cite{xmehta2008PRA} (see also Ref. \cite{xMacek}), 
the solutions of Eq.~(\ref{LS}) can be obtained by determining (for fixed values of $R$) the values of $s$ in 
which the determinant of the matrix
\begin{eqnarray}
Q=\left[\frac{3^{\frac{1}{4}}}{2^{\frac{1}{2}}R}\left(M^{(1)}+M^{(2)}P_-+M^{(3)}P_+\right)-1\right],
\end{eqnarray}
vanishes. Here, the matrices $(P_+)_{\Sigma\Sigma'}=\langle\Sigma|P_{123}|\Sigma'\rangle$ and $(P_-)_{\Sigma\Sigma'}=\langle\Sigma|P_{132}|\Sigma'\rangle$
represent, respectively, cyclic and anti-cyclic permutations of the three-body spin basis, and
\begin{eqnarray}
M^{(i)}_{\Sigma\Sigma'}=
\begin{cases} 
A_{\Sigma\Sigma'}^{(i)}s\cot(s\pi/2), & i=1, \\
-A_{\Sigma\Sigma'}^{(i)}\frac{4\sin(s\pi/6)}{\sqrt{3}\sin(s\pi/2)}, & i=2,3,
\end{cases}
\end{eqnarray}
is the two-body scattering length matrix written in the three-body spin basis,
\begin{eqnarray}
A_{\Sigma\Sigma'}^{(i)}=\langle\Sigma|\hat A_{\rm 2b}^{(i)}|\Sigma'\rangle.
\end{eqnarray}

Up to this point, our formulation does not account for any symmetry property of the system. In order to select 
the desired spin symmetry, we simply ``project'' the matrix $Q$ above to the particular subspace we are interested.
Therefore, the solutions with proper symmetry can be obtained by solving (for fixed values of $R$)
\begin{eqnarray}
{\rm det}\left[S (Q)S^{T}\right]=0,
\end{eqnarray}
where
\begin{align}
(S)_{\Sigma\Sigma'}=\langle\Sigma|\hspace{-0.05in}\left(\sum_{F_{\rm 2b}}|F_{\rm 3b}M_{F_{\rm 3b}}(F_{\rm 2b})\rangle \langle F_{\rm 3b}M_{F_{\rm 3b}}(F_{\rm 2b})|\hspace{-0.025in}\right)\hspace{-0.05in}|\Sigma'\rangle.
\end{align}
For $f=1$ atoms, for instance, the $F_{\rm 3b}=1$ solutions are determined by solving the transcendental equation,
\begin{align}
&\frac{3^{\frac{1}{4}}(a_0+a_2)s\cot(\frac{\pi}{2}s)}{2^{\frac{1}{2}}R}
-\frac{3^{\frac{1}{2}}(a_0a_2)s^2\cot(\frac{\pi}{2}s)^2}{2R^2}\nonumber\\
&-\frac{2^{\frac{3}{2}}(2a_0+a_2)\sin(\frac{\pi}{6}s)}{3^{\frac{5}{4}}\sin(\frac{\pi}{2}s)R}
+\frac{2~a_0a_2s\cot(\frac{\pi}{2}s)\sin(\frac{\pi}{6}s)}{\sin(\frac{\pi}{2}s)R^2}\nonumber\\
&+\frac{16~a_0a_2\sin(\frac{\pi}{6}s)^2}{3^{\frac{1}{2}}\sin(\frac{\pi}{2}s)^2R^2}=1. \label{EqF3b=1}
\end{align}
(Imaginary roots can be obtained by mapping $s\rightarrow is$.)
As we can see, the above transcendental equation depends on both two-body scattering lengths, $a_{0}$ and $a_2$. On the other hand,
the $F_{\rm 3b}=2$ and 3 solutions are obtained, respectively, through,
\begin{eqnarray}
\frac{3^{1/4}a_2s\cot(\frac{\pi}{2}s)}{2^{1/2}R}
+\frac{2~2^{1/2}a_2s\sin(\frac{\pi}{6}s)}{3^{1/4}\sin(\frac{\pi}{2}s)R}=1,\label{EqF3b=2}
\end{eqnarray}
and
\begin{eqnarray}
\frac{3^{1/4}a_2s\cot(\frac{\pi}{2}s)}{2^{1/2}R}
+\frac{4~2^{1/2}a_2s\sin(\frac{\pi}{6}s)}{3^{1/4}\sin(\frac{\pi}{2}s)R}=1, \label{EqF3b=3}
 \end{eqnarray}
which depend only on $a_{2}$. In Table \ref{TabRoots2} we list the solutions of Eqs.~(\ref{EqF3b=1})--(\ref{EqF3b=3}) for the regions
in $R$ in which the $s$ is constant. The values of $s$ listed represent the imaginary roots and/or the lowest real root.
In Table \ref{TabRoots2} we also list the corresponding values of $s$ relevant for $f=2$ spinor condensates. Note that
$F_{\rm 3b}=0$ and $1$ the possible values for $s$ depend only on $a_2$, while for $F_{\rm 3b}=2$ they depend on $a_0$, $a_2$ and $a_4$, $F_{\rm 3b}=3$ and $4$ on 
$a_2$ and $a_4$, and $F_{\rm 3b}=5$ and $6$ only on $a_4$.

\begin{table*}[htbp]
\caption{Values of $s_{\nu}$ relevant for $f$$=$$1$ and $2$ spinor condensates covering all possible regions of $R$  and for different magnitudes of the relevant 
scattering lengths. We list the lowest few values of $s_{\nu}$ for
each $F_{\rm 3b}$ and their multiplicity (superscript) if greater than one.
}\label{TabRoots2}
\begin{ruledtabular}
\begin{tabular}{lccccccc}
  {($f=1$)}  & $F_{\rm 3b}=1$ &$ F_{\rm 3b}=2$ & $F_{\rm 3b}=3$ \\
[0.05in]\hline
$R$$\ll$$|a_{\{0,2\}}|$  & $1.0062i,2.1662$ & $2.1662$ & $1.0062i,4.4653$ \\
$|a_0|$$\ll$$R$$\ll$$|a_2|$   & $0.7429$  & $2.1662$ & $1.0062i,4.4653$ \\
 $|a_2|$$\ll$$R$$\ll$$|a_0|$  & $0.4097$  &  $4$ & $2$ \\
$R$$\gg$$|a_{\{0,2\}}|$  & $2$  &  $4$  & $2$ \\
 [0.05in]\hline
  {($f=2$)}  &$F_{\rm 3b}=0$ &$F_{\rm 3b}=1$ &$F_{\rm 3b}=2$ &$ F_{\rm 3b}=3$ &$F_{\rm 3b}=4$ &$F_{\rm 3b}=5$ &$F_{\rm 3b}=6$ \\
[0.05in]\hline
$R$$\ll$$|a_{\{0,2,4\}}|$                   & $1.0062i, 4.4653$ &$ 2.1662 $& $1.0062i,2.1662^{(2)}$ & $1.0062i, 2.1662$          & $1.0062i, 2.1662$ & $2.1662$ & $1.0062i, 4.4653$ \\
$|a_0|$$\ll$$R$$\ll$$|a_{\{2,4\}}|$ & $1.0062i, 4.4653 $& $2.1662$ & $0.49050$                       & $1.0062i,2.1662 $          & $1.0062i, 2.1662$  & $2.1662$ & $1.0062i, 4.4653$\\
$|a_2|$$\ll$$R$$\ll$$|a_{\{0,4\}}|$ & $2$                           & $4$           & $0.7473i, 2.1662$          & $1.1044$                          & $0.66080$                     & $2.1662$ & $1.0062i, 4.4653$\\
$|a_4|$$\ll$$R$$\ll$$|a_{\{0,2\}}|$ & $1.0062i, 4.4653$ & $2.1662$ & $0.3788i, 2.1662$          & $0.5528i,3.5151$           & $0.52186$                    & $4$ & $2$\\
$|a_{\{0,2\}}|$$\ll$$R$$\ll$$|a_4|$ & $2$                           & $4$           & $0.97895$                       & $1.1044$                          & $0.66080$                     & $2.1662$ & $1.0062i, 4.4653$\\
$|a_{\{0,4\}}|$$\ll$$R$$\ll$$|a_2|$ & $1.0062i, 4.4653$ & $2.1662$ & $1.3173$                         & $0.5528i, 3.5151$           & $0.52186$                    & $4$ & $2$\\
$|a_{\{2,4\}}|$$\ll$$R$$\ll$$|a_0|$ & $2$                           & $4$           & $0.68609$                       & $2$                                     & $2$                   & $4$ & $2$\\
$R$$\gg$$|a_{\{0,2,4\}}|$                & $2$                           & $4$          &          $2$                                  & $2$                                           &$2$                        &$4$   &$2$
\end{tabular}
\end{ruledtabular}
\end{table*}

\section{Three-body scattering length operator}

In this section we will derive the form of the three-body scattering operator in which spin-exchange processes
are more evident and in which it is adequate for studies of three-body effects in the mean-field approach. We follow closely
the derivation of the corresponding two-body scattering length matrix established in Ref.~\cite{xSpinorTheo1}.
We start by expressing $\hat A_{\rm 3b}$ [Eq.~(7) of the main text] in terms of the projector operators for
each value of $F_{\rm 3b}$, ${\cal P}_{F_{\rm 3b}}$, as
\begin{eqnarray}
\hat{A}_{\rm 3b}&=&
\hspace{-0.1in}
\sum_{F_{\rm 3b}M_{F_{\rm 3b}}\atop F_{\rm 2b}}
\hspace{-0.1in}
a_{\rm 3b}^{(F_{\rm 3b})}|F_{\rm 3b}M_{F_{\rm 3b}}(F_{\rm 2b})\rangle \langle F_{\rm 3b}M_{F_{\rm 3b}}(F_{\rm 2b})|\nonumber \\
&=&
\sum_{F_{\rm 3b}}
a_{\rm 3b}^{(F_{\rm 3b})}{\cal P}_{F_{\rm 3b}},\label{SA3b}
\end{eqnarray}
where 
\begin{eqnarray}
{\cal P}_{F_{\rm 3b}}=\sum_{M_{F_{\rm 3b}}F_{\rm 2b}}
|F_{\rm 3b}M_{F_{\rm 3b}}(F_{\rm 2b})\rangle \langle F_{\rm 3b}M_{F_{\rm 3b}}(F_{\rm 2b})|.
\end{eqnarray}
Here, since the three-body states form a complete set of orthonormal states, the projection
operators satisfy the following
\begin{eqnarray}
\sum_{F_{\rm 3b}}{\cal P}_{F_{\rm 3b}}=1.\label{Compl}
\end{eqnarray}

Spin-exchange terms are determined by expressing them in terms of the projection operators. Using Eq.~(\ref{Compl}) and
the usual angular momentum relations, we have
\begin{align}
&\sum_{i<j}{\vec{f}_i\cdot\vec{f}_j}=\frac{1}{2}\left(\vec{F}_{\rm 3b}^2-\sum_{k}\vec{f}_k^2\right)\nonumber\\
&~~~~~~=\frac{1}{2}\left(\vec{F}_{\rm 3b}^2-\sum_{k}\vec{f}_k^2\right) \sum_{F_{\rm 3b}}{\cal P}_{F_{\rm 3b}}\nonumber\\
&~~~~~~=\sum_{F_{\rm 3b}}\frac{1}{2}\left[{F}_{\rm 3b}({F}_{\rm 3b}+1)-3f(f+1)\right] {\cal P}_{F_{\rm 3b}},
\end{align}
where $\vec{f}_i$ ($i$=1, 2 and 3) is the atomic hyperfine angular momentum for the atom 
$i$. For $f=1$ atoms, the above equation gives
\begin{align}
\sum_{i<j}{\vec{f}_i\cdot\vec{f}_j}=3{\cal P}_{3}-2{\cal P}_{1}-3{\cal P}_0,
\end{align}
and using ${\cal P}_3=1-{\cal P}_0-{\cal P}_1-{\cal P}_2$ from Eq.~(\ref{Compl}) we have
\begin{align}
{\cal P}_1=\frac{3-\sum_{i<j}{\vec{f}_i\cdot\vec{f}_j}-3{\cal P}_2-6{\cal P}_0}{5}.\label{P1}
\end{align}
Now, using Eqs.~(\ref{Compl}) and (\ref{P1}), we can express $\hat{A}_{\rm 3b}$ in Eq.~(\ref{SA3b}) as
\begin{eqnarray}
\hat{A}_{\rm 3b}=\alpha_{\rm 3b}+\alpha_{\rm 3b}^{\rm ex} \sum_{i<j}{\vec{f}_i\cdot\vec{f}_j},\label{SA3bS}
\end{eqnarray}
where the direct and spin-exchange terms are given by
\begin{eqnarray}
\alpha_{\rm 3b}=\frac{3a_{\rm 3b}^{(1)}+2a_{\rm 3b}^{(3)}}{5}~\mbox{and}~
\alpha_{\rm 3b}^{\rm ex}=\frac{a_{\rm 3b}^{(3)}-a_{\rm 3b}^{(1)}}{5}.
\end{eqnarray}
Note that,
since $F_{\rm 3b}=0$ and $2$ interactions are suppressed (see discussion in Section \ref{SecI}), 
we neglected in Eq.~(\ref{SA3bS}) the projection on the corresponding subspaces by setting ${\cal P}_0=0$ and ${\cal P}_2=0$.
We also note that, similar to the two-body case, the three-body spin-exchange term only describes processes in which 
$M_{\rm 3b}$ is conserved, i.e., $|m_{f_1}m_{f_2}m_{f_3}\rangle$=$|000\rangle$$\leftrightarrow$$|\pm10\mp1\rangle$ or
$|\pm100\rangle$$\leftrightarrow$$|\pm1\pm1\mp1\rangle$, as well as cyclic permutations of 
$|m_{f_1}m_{f_2}m_{f_3}\rangle$.

\begin{table}[htbp]
\caption{
Scattering length dependence for $F_{\rm 3b}=1$, 2 and 3 collision processes relevant for $f$=$1$ atom-dimer spinor condensates. Here, 
$K_{\rm ad}^{(0/d)}$ and $K_{\rm ad}^{(2/d)}$ are the decay rate of weakly bound $F_{\rm 2b}=0$ and 2 dimers into deeply bound states. Whenever $a_0>0$ and $a_2>0$,
$K_{\rm ad}^{(0/2)}$ ($K_{\rm ad}^{(2/0)}$) gives the decay rate for $F_{\rm 2b}=0$ ($F_{\rm 2b}=2$) dimers into $F_{\rm 2b}=2$ ($F_{\rm 2b}=0$) dimers due to collisions 
of atoms in states $|m_f\rangle$. $a_{\rm ad}^{(F_{\rm 2b})}$ is the atom-dimer scattering length. 
Here, $k_{\rm ad}^2=2\mu_{\rm ad}E_{\rm col}$, $M(a)$, $P(a)$, $T(a)$ and $O(a)$ are given in Eqs.~(\ref{SMformula})-(\ref{SOformula}),
$s_1\approx0.7429$ ($|a_2|$$\gg$$|a_0|$) and $s_1\approx0.4097$ ($|a_0|$$\gg$$|a_2|$) for $F_{\rm 3b}$$=$$1$ and
$s_1\approx2.1662$ for $F_{\rm 3b}$$=$$2$.}\label{TabScatAD}
\begin{ruledtabular}
\begin{tabular}{cccc}
$(F_{\rm 3b}=1)$& $a_2\gg a_0$  & $a_2\gg |a_0|$  & $|a_2|\gg a_0$ \\
[0.05in] \hline 
$K_{\rm ad}^{(0/d)}/a_0$ & {\small$P_{s_0}^{\eta}(a_0)$} & --- & {\small$P_{s_0}^{\eta}(a_0)$}  \\
$K_{\rm ad}^{(2/d)}/a_2$ & {\small $\left(\frac{a_0}{a_2}\right)^{2s_1}$} & {\small$P_{s_0}^{\eta}(a_0) \left(\frac{a_0}{a_2}\right)^{2s_1}$} & ---  \\
$K_{\rm ad}^{(2/0)}/a_2$ & {\small $M_{s_0}^{\eta}(a_0) \left(\frac{a_0}{a_2}\right)^{2s_1}$} & --- & --- \\
$K_{\rm ad}^{(0/2)}/a_2$ & {\small $K_{\rm ad}^{(2/0)}\left(\frac{a_{0}}{a_2}\right) k_{\rm ad}$} & --- & --- \\
$a_{\rm ad}^{(0)}/a_0$ & {\small$ T_{s_0}^{\eta}(a_0)$} & --- & {\small$ T_{s_0}^{\eta}(a_0)$}  \\
$a_{\rm ad}^{(2)}/a_2$ & {\small$1$$+$$O_{s_0}^{\eta}(a_0)\left(\frac{a_0}{a_2}\right)^{4s_1}$} & {\small$1$$+$$T_{s_0}^{\eta}(a_0)\left(\frac{a_0}{a_2}\right)^{4s_1}$} & --- \\
[0.05in] \hline
$(F_{\rm 3b}=1)$& $a_0\gg a_2$  & $a_0\gg |a_2|$  & $|a_0|\gg a_2$ \\
[0.05in] \hline 
$K_{\rm ad}^{(0/d)}/a_0$ & {\small $\left(\frac{a_2}{a_0}\right)^{2s_1}$} & {\small$P_{s_0}^{\eta}(a_2) \left(\frac{a_2}{a_0}\right)^{2s_1}$} & ---  \\
$K_{\rm ad}^{(2/d)}/a_2$ & {\small$P_{s_0}^{\eta}(a_2)$} & --- & {\small$P_{s_0}^{\eta}(a_2)$}  \\
$K_{\rm ad}^{(0/2)}/a_2$ & {\small $M_{s_0}^{\eta}(a_2) \left(\frac{a_2}{a_0}\right)^{2s_1}$} & --- & --- \\
$K_{\rm ad}^{(2/0)}/a_2$ & {\small $K_{\rm ad}^{(0/2)}\left(\frac{a_2}{a_0}\right) k_{\rm ad}$} & --- & --- \\
$a_{\rm ad}^{(0)}/a_0$ & {\small$1$$+$$O_{s_0}^{\eta}(a_2)\left(\frac{a_2}{a_0}\right)^{4s_1}$} & {\small$1$$+$$T_{s_0}^{\eta}(a_2)\left(\frac{a_2}{a_0}\right)^{4s_1}$} & --- \\
$a_{\rm ad}^{(2)}/a_2$ & {\small$ T_{s_0}^{\eta}(a_2)$} & --- & {\small$ T_{s_0}^{\eta}(a_2)$}  \\
[0.05in] \hline
$(F_{\rm 3b}=2)$& $a_2\gg r_0$ & $(F_{\rm 3b}=3)$ & $a_2\gg r_0$   \\
[0.05in] \hline
$K_{\rm ad}^{(2)}/a_{2}$ & {\small$\left(\frac{r_{\rm vdW}}{a_2}\right)^{2s_1}$} & $K_{\rm ad}^{(2)}/a_{2}$ & {\small$P_{s_0}^{\eta}(a_2)$} \\
$a_{\rm ad}^{(2)}/a_{2}$ & {\small$1$$+$$\left(\frac{r_{\rm vdW}}{a_2}\right)^{4s_1}$} & $a_{\rm ad}^{(2)}/a_{2}$ & {\small$T_{s_0}^{\eta}(a_2)$}  \\
\end{tabular}
\end{ruledtabular}
\end{table}

\section{Analytical Formulas For Atom-dimer collisions}

In deriving the scattering length dependence of atom-dimer collisional processes, the influence of Efimov physics in inelastic
processes generates interference and resonant effects which are parametrized, respectively, by 
\begin{align}
&M_{s_0}^{\eta}(a)=\alpha~{e^{-2\eta}}\left[\sin^2\left(\mbox{$|s_{0}|\ln \frac{a}{r_{\phi}}$}\right)+\sinh^2\eta\right],\label{SMformula}\\ 
&P_{s_0}^{\eta}(a)=\beta~\frac{\sinh2\eta}{\sin^2\left(|s_{0}|\ln \frac{a}{r_{\phi}}\right)+\sinh^2\eta}\label{SPformula},
\end{align}
where $\alpha$ and $\beta$ are universal constants and $r_{\phi}$$=$$r_{\rm vdW}e^{-\phi/|s_0|}$ is the three-body parameter, incorporating the three-body short-range physics through the 
phase $\phi$ \cite{xJILA_Theo}. $\eta$ is the three-body inelasticity parameter \cite{xbraaten2006PR,xwang2013AAMOP} which encapsulates the probability for decay into deeply bound molecular
states. Similarly, Efimov physics also manifest in atom-dimer elastic processes through resonant and interference effects parametrized by
\begin{align}
&T_{s_0}^{\eta}(a)=\alpha+\beta~\frac{2\sin\left(|s_{0}|\ln \frac{a}{r_{\phi}}\right)\cos\left(|s_{0}|\ln \frac{a}{r_{\phi}}\right)}{\sin^2\left(|s_{0}|\ln \frac{a}{r_{\phi}}\right)+\sinh^2\eta},\label{STformula}\\
&O_{s_0}^{\eta}(a)=\alpha+\beta{e^{-2\eta}}\left[\sin^2\left(\mbox{$|s_{0}|\ln \frac{a}{r_{\phi}}$}\right)+\sinh^2\eta\right].\label{SOformula}
\end{align}
In Table \ref{TabScatAD}, we summarize the atom-dimer decay rates and elastic parameters for $f=1$ atoms.

\end{document}